\documentclass[aip,jcp,amsmath,amsymb,reprint]{revtex4-1}
\usepackage{graphicx}
\usepackage{bm}

\begin{document}

	\title{Designing convex repulsive pair potentials that favor assembly of kagome and snub square lattices}

	\author{William D. Pi\~neros} 
	\affiliation{Department of Chemistry, University of Texas at Austin, Austin, TX 78712}
	\author{Michael Baldea} 
	\author{Thomas M. Truskett}
	\affiliation{McKetta Department of Chemical Engineering, University of Texas at Austin, Austin, TX 78712}
	\date{\today}
	\begin{abstract}
		Building on a recently introduced inverse strategy, isotropic and convex repulsive pair potentials were designed that favor assembly of particles into kagome and equilateral snub square lattices. The former interactions were obtained by numerical solution of a variational problem that maximizes the range of density for which the ground state of the potential is the kagome lattice. Similar optimizations targeting the snub square lattice were also carried out, employing a constraint that required a minimum chemical potential advantage of the target over select competing structures. This constraint helped discover isotropic interactions that meaningfully favored the snub square lattice as the ground state structure despite the asymmetric spatial distribution of particles in its coordination shells and the presence of tightly competing structures. Consistent with earlier published results [Pi\~neros et al. J. Chem. Phys. 144, 084502 (2016)], enforcement of greater chemical potential advantages for the target lattice in the interaction optimization led to assemblies with enhanced thermal stability.
	\end{abstract}

	\maketitle 

\section{Introduction}
	 Fabricating materials with predetermined structural features, especially at the nanoscale, remains an outstanding challenge. While technological advances in top-down approaches such as lithography have expanded the possibilities in this arena, these methods are often prohibitively slow and expensive to implement at the length scales of interest for industrial nanomanufacturing applications\cite{PhotonicMatsDesign,PhotonicMatsDesign2}. Bottom-up approaches such as the designed self assembly of particles (molecules, nanocrystals, colloids, etc) into desired morphologies represent promising alternatives \cite{SelfAssemblySuperlatticeCatalysis,JanusParticlesSelfAssemblyRev,SelfAssemblyForcesReview}. The success of these latter strategies relies upon the ability to ``tune'' the relevant interactions for assembly \cite{ColloidInteractionTuning_1,SelfAssemblyForcesReview,ColloidInteractionsReview}, e.g., by synthesizing nanoparticles with precisely defined shape, size, and surface chemistry, as well as by dispersing such particles into different solution environments. It also depends upon the ability to guide such efforts by determining in advance which types of interactions will deliver assemblies with the desired structural features at thermodynamic equilibrium. 
	 
	 Generally, methods to address the latter can be separated into {\em forward} approaches, wherein a series of candidate particle systems (or model interactions) are judiciously chosen and subsequently tested for suitability, or {\em inverse} methods, wherein interactions that could produce the desired structure are directly determined theoretically, typically via a constrained optimization calculation.  While forward strategies are often intuitive and straightforward to apply, they can be time consuming and expensive to successfully execute. Inverse design approaches, although not always as easy to formulate and solve, offer the possibility of a more precise and definitive determination of the interactions required to achieve a desired end result \cite{InvDesignTechRev,InvDesignGeneral,InvDesignPerspective}. 

 	 One classic example of an inverse materials design problem is determining which isotropic pair potentials stabilize a specified ground state lattice.  Previous studies addressing this type of problem found pair interactions that favor a wide variety of open two dimensional (e.g., square and honeycomb) \cite{InvDesignKagome,InvDesignKagomeFunctionalMethod,MT_SquareHoneyConvexCom,RT_HoneyDoubleWell} and three dimensional (e.g., diamond, wurtzite, and simple cubic) \cite{MT_DiamondConvex,InvDesignKagomeDiamond} target lattices, many of which readily assemble from the disordered fluid state upon cooling in a simulation. Interestingly, using stochastic optimization methods, Jain et al. have demonstrated that these low-coordinated ground state structures can even be stabilized over a wide range of density by convex repulsive pair potentials\cite{AvniPotModel,Avni3DLattices,AvniDimTransfer}, qualitatively similar to the effective interactions \cite{SelfAssemblyForcesReview} characteristic of soft gels, micelles, star polymers, etc. The specific functional form of the interactions considered in these latter studies is given by a generalized form of the Fomin potential \cite{AvniPotModel}
	\begin{equation}
		\begin{aligned}
		\phi(r/\sigma) &= \epsilon \lbrace A(r/\sigma)^{-n} + \sum_{i=1}^{2} \lambda_i(1-\tanh[k_i(r/\sigma-\delta_i)]) \\ 
			       &+f_{\text{shift}}(r/\sigma) \rbrace H[(r_{\text{cut}}-r)/\sigma]
		\label{eq:potform}
		\end{aligned}
	\end{equation}
where $\sigma$ and $\epsilon$ represent characteristic length and energy scales respectively; $H$ is the Heaviside function; $\{A,n,\lambda_i, k_i,\delta_i\}$ are variable parameters, one of which is fixed to ensure $\phi(1)=\epsilon$; $r_{\text{cut}}$ is a cut-off radius; $f_{\text{shift}}$ is a quadratic function $f_{\text{shift}}(r/\sigma)= P (r/\sigma)^2 + Q r/\sigma + R$ added to enforce $\phi(r_{\text{cut}}/\sigma)= \phi'(r_{\text{cut}}/\sigma)= \phi''(r_{\text{cut}}/\sigma)= 0$.  Pair potentials of the form provided by Eq.~\ref{eq:potform} are considered in this work as well. For notational simplicity, all quantities reported from here forward are implicitly nondimensionalized by appropriate combinations of $\epsilon$ and $\sigma$.

	Recently, we improved the efficiency of Jain et al.'s optimization strategy by reformulating it as an analytical nonlinear program that can be solved numerically \cite{SquLat_dmu_opt}. Using this approach, we studied the consequences of designing interactions that stabilize the target structure over a wide density range while requiring that it maintain a minimum chemical potential advantage $\Delta \mu$ over select competing lattices.  Interestingly, this chemical potential constraint led to the discovery of new interactions that increased target structure thermal stability (albeit at the expense of stability with respect to changes in density).
	
	Here, we build on this optimization framework and consider two new and contrasting design targets for self assembly from particles with convex repulsive potentials: the kagome and the equilateral snub square lattices. While both target structures have been previously demonstrated to self-assemble using patchy particles\cite{ArchStructures_PatchyParticles}, polygons \cite{ArchStructures_PatchyPolygons} or binary mixtures \cite{ArchStructures_BinaryMix}, this work is the first to address them using isotropic convex repulsive interactions in a single component system. In particular, assembly of the kagome lattice is of interest in magnetic materials due to its unusual properties arising from geometrical frustration \cite{KagomeMagFrustationRev,TopologicalExcitationsKagomeMagnet}.  Materials exhibiting a kagome lattice are known to be difficult to synthesize experimentally, and though there are now a few isotropic model potentials known to stabilize this structure \cite{InvDesignKagome,InvDesignKagomeFunctionalMethod,BT_QuinticPotentialKagomeHoneycomb,ZT_MuOptKagomeAsymLats}, none of them are of the simple convex repulsive type considered here. Additionally, the kagome lattice presents an attractive design target for an isotropic potential in one respect: the spatial distributions of particles in its coordination shells are symmetric. In fact, we show here that one can design convex repulsive interactions that stabilize the kagome lattice ground state over a wide range of density, and that particles with these interactions readily self-assemble into the kagome lattice from the fluid upon cooling.

	The equilateral snub square lattice, on the other hand, presents a significantly more challenging design target for an isotropic pair potential. First, the neighbors in the coordination shells surrounding each particle in this lattice have an asymmetric spatial distribution. The difficulties in stabilizing such asymmetric arrangements with an isotropic potential have been discussed previously in an insightful paper by Zhang et al., \cite{ZT_MuOptKagomeAsymLats} and, to our knowledge, have yet to be overcome in a design application with an isotropic, convex repulsive pair potential. The second complication is that the snub square lattice is very similar to the elongated triangular lattice, sharing the same specific area at close packing and identical numbers of neighbors in the first two coordination shells (reminiscent of diamond and wurtzite lattices in three dimensions). Finally, the highly coordinated first shell of the snub square lattice (with five neighbors) also puts it in close competition with the common triangular lattice (with six neighbors). In this work, to successfully overcome these hurdles in designing interactions that assemble particles into the snub square lattice, we enforce a minimum chemical potential advantage of the target structure over select competing lattices during the potential optimization. We consider a weak and a strong constraint and find that, consistent to previous work, the latter leads to enhanced thermal stability of the designed snub square structure \cite{SquLat_dmu_opt}.
	
	The article is organized as follows. In Section \ref{sec:Methods}, we briefly describe our previously introduced method for designing isotropic and convex repulsive pair potentials to stabilize a prescribed lattice, extended here to include the range of the potential as a decision variable. We also describe Monte Carlo simulation quenches that we use to test whether the particles interacting via the designed potentials readily assemble into the target structures rapid isochoric cooling from the fluid state. Next we present the designed pair potentials for the kagome and the snub square lattices in Section \ref{sec:Results} as well as the results from the Monte Carlo quenches. In the case of the snub square lattice, we explore the thermal stability of structures associated with two pair potentials designed using different constraints on their chemical potential advantage over select competitors. Finally, we present concluding remarks in Section \ref{sec:Conclusion}.

\section{Methods} 
\label{sec:Methods}
	\subsection{Inverse Design of the Pair Potential}
	The inverse design optimization we use here is formulated analytically as a nonlinear program in which one seeks parameters of the pair potential of Eq.~$1$ that (1) make it convex repulsive and (2) maximize the range of density for which the target structure is the ground state (i.e., has lower chemical potential than equi-pressure competing lattices). This optimization can be cast in a way that also includes a constraint that ensures that the target structure, at a given density $\rho_0$, exhibits a minimum prescribed chemical potential advantage $\Delta \mu$ over those lattices in a select pool of `flag-point' competing structures, a strategy shown previously to enhance the target's thermal stability$^{20}$. Additionally, this analytical formulation offers solution strength and computing time advantages that are more difficult to achieve with familiar stochastic approaches (e.g. simulated annealing, genetic algorithms, etc). The complete formulation and implementation of this approach, including the required equations and their numerical solution using GAMS (Generalized Algebraic Modeling System) \cite{GAMSWorldBank,GamsSoftware2013,GamsGuide2013}, are presented elsewhere \cite{SquLat_dmu_opt}. 

In previous work$^{20}$ , we considered the pair potential cut-off, $r_{\text {cut}}$, to be a fixed parameter. As such, we constrained the second derivative of the potential using the inequality
	\begin{equation}
		\phi''(\mathbf{r}) > 0 
	\label{eq:d2phi_fix}	
	\end{equation} 
where $\mathbf{r}$ was discretized as a fixed distribution of pair separations between $0$ and $r_{\text {cut}}$.  Here, we allow for $r_{\text{cut}}$ to itself be a decision variable in the optimization that can vary between $r_{\text{c,min}}$ and $r_{\text{c,max}}$. Including $r_{\text{cut}}$ as a decision variable allows the optimizer to use an additional degree of freedom to adjust near-feasible solutions to conform to our interaction constraints (i.e., normality and convexity).  Additionally, having a $r_{\text{cut}}$ as a decision variable ensures that the optimization has the flexibility to extend the pair interaction to include sufficient coordination shells to stabilize a desired target. Considering these advantages, we restrict Eq.~\ref{eq:d2phi_fix} to apply between $0$ and $r_{\text{c,min}}$, and then we add an equation,
	\begin{equation}
		\phi''(\mathbf{r}_v) > 0, 
	\end{equation}
where $\mathbf{r}_v$ is represented by ten points evenly distributed between $r_{\text{c,min}}$ and $r_{\text{c,max}}$. More specifically, $\mathbf{r}$ in eq. \ref{eq:d2phi_fix} is evaluated at fifty points between 0 and $r_{\text{c,min}}$ distributed in an approximately 1:6:5 ratio from ranges $[0.3,0.8)$, $[0.8,1.2)$, and $[1.2,r_{\text{c,min}})$ (modified slightly as to best fit individual targets). This generalization reduces to the originally considered sixty fixed points in $\mathbf{r}$ if $r_{\text{cut}}$ is assigned a constant value. For optimizations targeting the kagome lattice with no $\Delta\mu$ constraint and the snub square lattice with the weak constraint $\Delta\mu=0.01$, we assigned $r_{\text{cut}}$ to constant values of 3 and 2, respectively. For the optimization targeting the snub square lattice with the strong constraint $\Delta\mu=0.04$, we employed a variable $r_{\text{ cut}}$ from $r_{\text{c, min}}=1.8$ to $r_{\text{c, max}}=2.5$ and obtained an optimized value of $r_{\text{cut}}=1.80082033$.  

	\subsection{Competing Lattices}
	\label{sec:CompLattices}
	In carrying out pair potential optimizations, one compares the chemical potential of the target lattice to that of lattices in a small pool of competing structures. For our initial optimization, this competitive pool comprised Bravais and non-Bravais lattices which commonly occur in the phase diagram of two-dimensional systems with soft, repulsive interactions \cite{SquLat_dmu_opt}. Once an optimized pair potential was obtained considering this initial competitive pool, a forward calculation was carried out to determine its ground-state phase diagram. Any new structures that appear in this phase diagram with chemical potentials comparable to the target lattice were added to the competitive pool, and a new optimization was performed. This procedure was repeated until no new competitive lattices were discovered in the forward calculation of the ground-state phase diagram for the optimized potential.

	In order to design interactions with enhanced thermal stability of the target structure, potential optimizations can also be carried out with a constraint that enforces a minimum chemical potential advantage $\Delta \mu$ of the target at density $\rho_0$ over a small pool of equi-pressure lattices that serve as `flag points' on the $\mu$ hypersurface \cite{SquLat_dmu_opt}. Ideally, one chooses flag-point lattices to be natural competing structures, some of which may be related to the target by a simple disturbance of the former's ideal configuration. More information and examples on how the choice of flag-point lattices for a given target might be determined are provided in our previous work \cite{SquLat_dmu_opt}.

	For the present study, the final competing pools for different target structures are as follows, where an asterisk indicates a lattice is also used as a flag-point competitor for the chemical potential constraint. 	For the kagome target with no explicit $\Delta\mu$ constraint, the resulting competing lattices included elongated triangular, triangular, equilateral snub square, honeycomb, rectangular ($b/a=2.01$), `kagome-B' (i.e. a kagome with non-uniform aspect ratio), and distorted-honeycomb. Interestingly, twisted kagome, a related kagome structure by rigid rotation of the triangular motifs, did not arise explicitly as a competitor in the forward calculations but might prove necessary as a flag-point competitor in a $\Delta\mu$ constraint optimization. For the equiliateral snub square target for which a weak ($\Delta\mu=0.01$) and strong ($\Delta\mu=0.04$) chemical potential advantage constraint was enforced, the competing lattices included elongated triangular*, triangular*, honeycomb*, square*, elongated triangular* ($b/a=1.15$), rectangular ($b/a=1.17$), snub square ($b/a=1.005$) and distorted snub square. For full details on kagome-B, distorted-honeycomb and distorted snub square lattices, please see the Appendix.  

	\subsection{Monte Carlo Quenches}
	The assembly of target structures from the optimized potentials was tested via quenches from sufficiently high temperature fluid states in canonical Monte Carlo simulations utilizing periodic boundary conditions.
	
	For the kagome lattice, a system of $N=1200$ particles and a simulation cell with dimensions adjusted to fix density at $\rho=1.40$ were first equilibrated in the fluid state at $T=0.023$ and then isochorically quenched in two steps: first to $T=0.01$ and subsequently to $T=0.005$. For the equilateral snub square forming potentials, similarly sized systems became kinetically trapped in defective structures during isochoric Monte Carlo quenches from the fluid, but smaller systems readily assembled into the designed target structure. For this lattice, we report results with $N=64$ particles with simulation cell size adjusted to fix density to $\rho=1.425$. The pair potential designed using the chemical potential constraint $\Delta\mu=0.01$ could be equilibrated in the fluid state at $T=0.0655$, and it was subsequently quenched to $T=0.0309$ to induce assembly of the snub square lattice.  The potential designed using the chemical potential constraint $\Delta\mu=0.04$ exhibited a snub square structure with enhanced thermal stability, and thus the equilibration of the fluid state was carried out at the higher temperature of $T=0.1$.  For this system, assembly from the fluid state was observed in a two step quench, first cooling to $T=0.482$ and then to $T=0.0309$ to refine the structure.
	
	\begin{figure*}[ht]
	\includegraphics[scale=0.85]{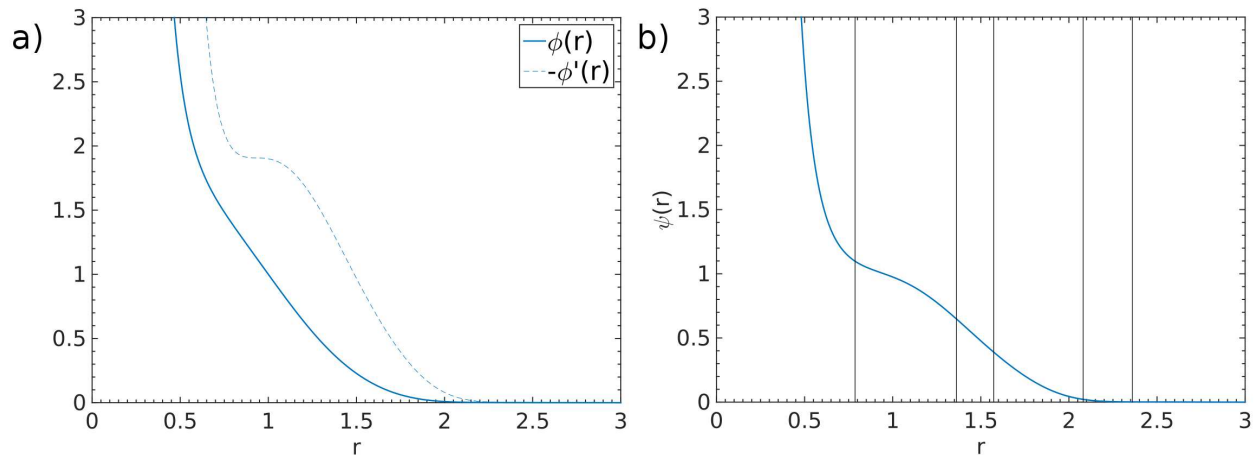}
	\caption{a) Pair potential $\phi(r)$ and force $-\phi'(r)$ of Eq.~\ref{eq:potform} with parameters optimized to maximize the range of density for which the ground state is the kagome lattice. b) Plot of $\psi(r)$ of Eq.~4 with kagome lattice coordination shell distances indicated by vertical black lines located for the optimization density $\rho=1.40$.}
	\label{fig:kag_potpsi}
	\end{figure*} 
\section{Results and Discussion}
\label{sec:Results}
	\begin{table*}[ht]
	\caption{Parameters for the convex repulsive pair potential $\phi(r)$ of Eq.~\ref{eq:potform} found to maximize the density range for which kagome and equilateral snub square lattices are the ground states. SS-A and B refer to parameters for two snub square favoring potentials optimized with $\Delta\mu$ constraint values of $0.01$ and $0.04$, respectively.}
	\label{tab:potpars}
	\begin{ruledtabular}
	\begin{tabular}{lccccccccc}
			&$A$		&$n$		&$l_1$		&$k_1$		&$d_1$		&$l_2$		&$k_2$		&$d_2$		&$r_c$ \\ \hline
	kagome		&0.01978	&5.49978	&-0.06066	&2.53278	&1.94071	&1.06271	&1.73321	&1.04372	&3.00000 \\
	SS-A 		&2.55737	&1.53719	&0.10022	&6.24964	&1.48785	&0.15066	&7.72221	&1.12084	&2.00000 \\	
	SS-B		&26.26595	&1.75476	&1.00266	&3.41639	&1.52736	&-13.85332	&3.92417	&0.68249	&1.80082 \\
	\end{tabular}
	\end{ruledtabular}
	\end{table*}
Using the optimization strategy described in Section~\ref{sec:Methods}, we were able to find a convex repulsive potential $\phi(r)$ that maximized the density range $\Delta\rho$ for which the kagome lattice was the ground state with a value of $\Delta\rho=0.415$. In one sense, this may not seem surprising. As alluded to in the Introduction, structures such as the kagome lattice with symmetric spatial distributions of particles in their coordination shells tend to be more amenable to stabilization with an isotropic pair potential. This favorable predisposition, however, does not guarantee the existence of a desired interaction form. In fact, Eq.~\ref{eq:potform} with optimized parameters provided in Table~\ref{tab:potpars} is, to our knowledge, the only purely convex repulsive pair interaction reported to stabilize the kagome lattice, and our ability to easily find it is a testament to the robustness of the design approach. The other pair potentials known to stabilize this structure do so via incorporation of other features at specific separations (e.g., attractive wells, concave shoulders, etc.) \cite{InvDesignKagome,InvDesignKagomeFunctionalMethod,BT_QuinticPotentialKagomeHoneycomb,ZT_MuOptKagomeAsymLats}, which may be more challenging to realize in practice.

How can a seemingly featureless convex repulsive pair interaction select a structure as specific as the kagome lattice over its competitors?  As can be seen in Fig.~\ref{fig:kag_potpsi}a), the principal features of the designed kagome potential lie within the radial range of $0.5 \lesssim r \lesssim 2$, where it transits from a core ($r \lesssim 0.5$) into an approximately linear ramp (i.e. the force, $-\phi(r)'$ is nearly constant value at these points as shown by the dashed line in the figure). The implications of this form can be more clearly appreciated by considering a function $\psi(r)$ which we previously showed\cite{SquLat_dmu_opt} is related not only to the pair potential but also to the chemical potential of a ground-state lattice: 
               \begin{align}
			\mu_l=\sum_{i}^{r_{i,l} < r_c} n_{i,l}\psi(r_{i,l}(\rho_l)) \\ \nonumber
                        \psi(r) \equiv \frac{\phi(r)}{2}  - \frac{r \phi'(r)}{4}
                \label{eq:psir}
                \end{align} 
where $r_{i,l}$ denotes the $i^{\text{th}}$ coordination shell distance for a lattice of type $l$ at density $\rho_l$. In short, $\psi(r)$ helps understand what radially-varying `weights' (due to the form of the pair potential) would multiply the occupation numbers $n_{i,l}$ in a given lattice $l$ to determine the coordination shell contributions to its chemical potential $\mu_l$.

	A plot of $\psi(r)$ for the kagome potential is shown in Fig.~\ref{fig:kag_potpsi}b) at a density ($\rho=1.4$) near the middle of the target lattice's stable range on the ground-state phase diagram. Perhaps the most prominent feature of this function is the shoulder that it exhibits for separations in the range $0.6 \lesssim r \lesssim 1.35$ before decaying to close zero by $r \sim 2.2$.  In particular, this shoulder shape helps penalize the heavily coordinated first shell of competitors such as triangular and snub square lattice (6 and 5 neighbors respectively) throughout the kagome lattice density range by keeping the relative contribution from this shell nearly constant. Similarly, the decaying tail destabilizes more evenly spread shells from competitors like the rectangular lattice while making $\psi(r)$ small and smooth enough to diminish the contribution from the more heavily coordinated third shell in the kagome lattice (6 neighbors). Together, these features help establish the kagome lattice as the stable structure and explain the wide density range achieved as per our design goal.
	\begin{figure}[ht]
	\includegraphics[scale=0.87]{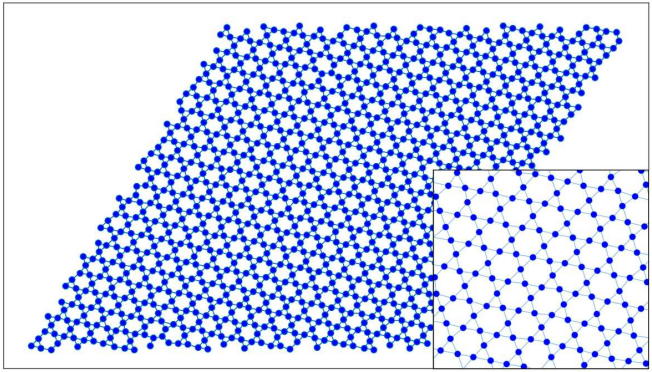}
	\caption{Configuration snapshot from a Monte Carlo simulation of a kagome lattice that self assembled from a fluid of the optimized potential (discussed in text) upon quenching to $T=0.005$ at $\rho=1.4$ Inset depicts a zoomed in view of a representative region. See Section ~\ref{sec:Methods} for simulation details.} 
	\label{fig:kag_conf}
	\end{figure} 

	To verify thermal stability of the kagome lattice with the optimized potential, we carried out Monte Carlo simulations (as described in Section ~\ref{sec:Methods}) from the disordered fluid state for which the kagome lattice readily assembled upon quenching.  In Figure~\ref{fig:kag_conf}, we show a representative configuration of the quenched structure. As can be seen, only very minor imperfections are present in the assembled kagome lattice, arising from the usual misalignment of the nucleated crystal relative to the periodically replicated simulation cell.
	\begin{figure}[ht]
	\includegraphics[scale=0.27]{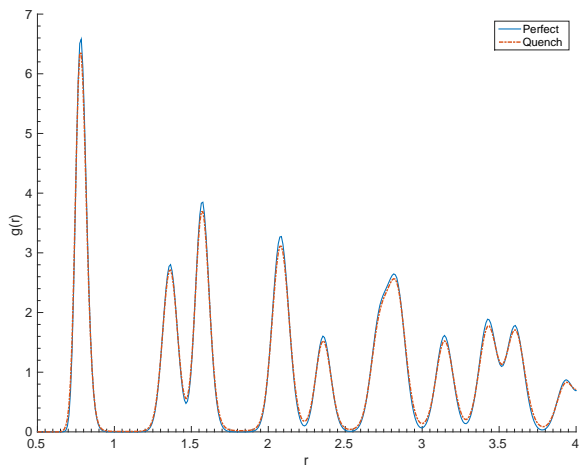}
	\caption{Radial distribution function of the perfect equilibrium kagome crystal (solid blue line) and the kagome crystal assembled from a fluid quench (dashed red) at $T=0.005$.}
	\label{fig:kag_gr}
	\end{figure} 
	 
	 We can more concretely quantify the order displayed by the assembled kagome lattice by comparing its radial distribution function $g(r)$ to that of an equilibrated perfect crystal at the same temperature. As shown in figure \ref{fig:kag_gr}, the quenched structure very nearly matches the perfect ordering in all coordination shells.  Further supporting this observation, the potential energy difference between the quenched-assembled and equilibrated perfect crystal structures was less than  $0.05\%$.  Overall, these results show that our designed convex repulsive pair potential thermodynamically favors the kagome lattice, a structure which readily self assembles from the fluid state upon cooling.  

	Moving on to our next design target, we present results for the significantly more challenging equilateral snub square lattice.  The inherent difficulty is discovering an isotropic pair potential, and especially a convex repulsive interaction, that can not only selectively lower the chemical potential of this target relative to similar triangular and elongated triangular lattices, but also stabilize the snub square structure despite the asymmetric spatial distribution of neighbors in its coordination shells.

	As described in Section~\ref{sec:Methods}, to help ensure a significant free energy gap relative to competitors, we enforced a nonzero chemical potential advantage $\Delta\mu$ constraint for the target lattice over appropriately selected `flag-point' lattices in the ground state. Additionally, considering previous results where the severity of such a $\Delta\mu$ constraint correlated with the thermal stability of the designed target structure\cite{SquLat_dmu_opt}, we explored the behaviors of assemblies that resulted from optimizations with weak and strong constraints ($\Delta\mu=0.01$ and $\Delta\mu=0.04$, respectively). The potential parameters obtained from this procedure are provided in Table \ref{tab:potpars}. To our knowledge, these are the only isotropic and convex-repulsive interactions that have been reported to stabilize the equilateral snub-square lattice.
	\begin{figure*}[ht]
	\includegraphics[scale=0.85]{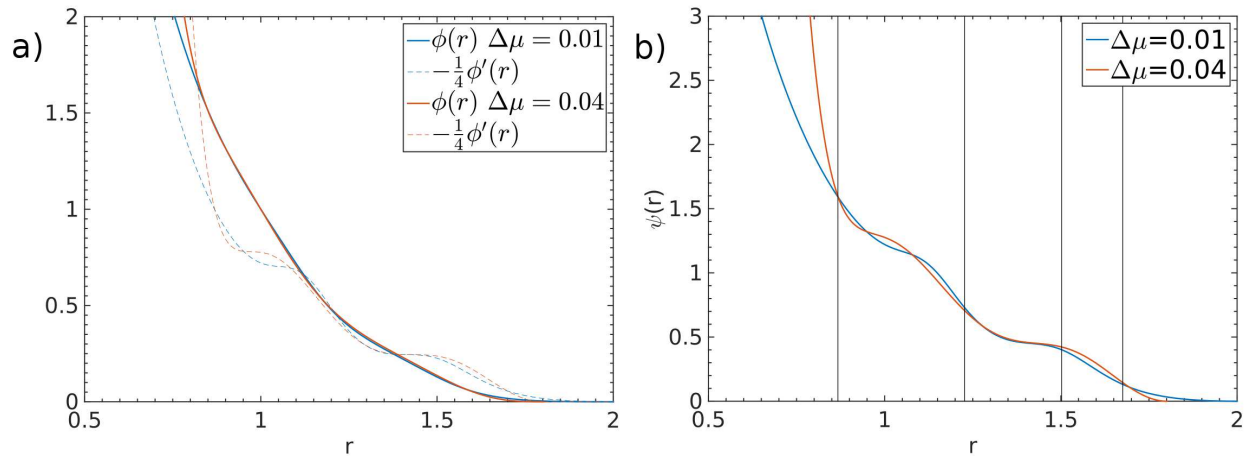}
	\caption{a)  Pair potential $\phi(r)$ and force $-\phi'(r)/4$ (divided by four here to fit graph) of Eq.~\ref{eq:potform} with parameters optimized to maximize the range of density for which the ground state is the equilateral snub square lattice, subject to a chemical potential advantage constraint $\Delta\mu=0.01$ (blue) and $\Delta\mu=0.04$ (red) of the target structure over select `flag-point' competitors.
	b) Plot of $\psi(r)$ of Eq.~4 for these pair potentials. Vertical black lines indicate snub-square lattice coordination shell distances at optimization density $\rho=1.425$.} 
	\label{fig:snub_potpsi}
	\end{figure*}

	Considering the difference in constraints used to obtain these potentials, it is insightful to consider plots of optimized $\phi(r)$ and the corresponding auxiliary function $\psi(r)$ as shown in Figure \ref{fig:snub_potpsi}. As can be seen in the potential plot (solid line), both pair potentials show a `core' ($r \lesssim 0.8$) and a `two-ramp' repulsion in the ranges $0.8 \lesssim 1.25$ and $1.25 \lesssim r \lesssim 1.65$ (i.e. near constant force--see dashed plots), respectively that together result in a two-plateau structure for $\psi(r)$. In analogy to the optimized kagome interaction, the first plateau is situated such that the target lattice's second coordination shell falls after the first plateau, ensuring that the first coordination shell is more strongly weighted than more distant shells in the chemical potential.  The second plateau also plays a critical role--It penalizes competing lattices relative to the snub square by creating a high value of $\psi(r)$ at distances corresponding to their third coordination shell, which raises the chemical potential of competitors with more third-shell neighbors (e.g., the triangular lattice has six neighbors in its third shell versus only one for the snub square lattice). Note also that the elevated $\Delta\mu=0.04$ constraint led to steeper radial decays in $\psi(r)$ near the core and potential cut-off regions as well as sharpened plateau regions in comparison to the potential obtained with $\Delta\mu=0.01$.\cite{AnalysisSupp} These trends are consistent with previous results on designing potentials to stabilize square lattices \cite{SquLat_dmu_opt}, where increasing the $\Delta\mu$ constraint led to similarly sharpened interaction features that more heavily penalized competing structures. Such features also led to enhanced target lattice thermal stability, which is consistent with our current results as we discuss next.  

	\begin{figure*}[ht]
	\includegraphics[scale=0.4]{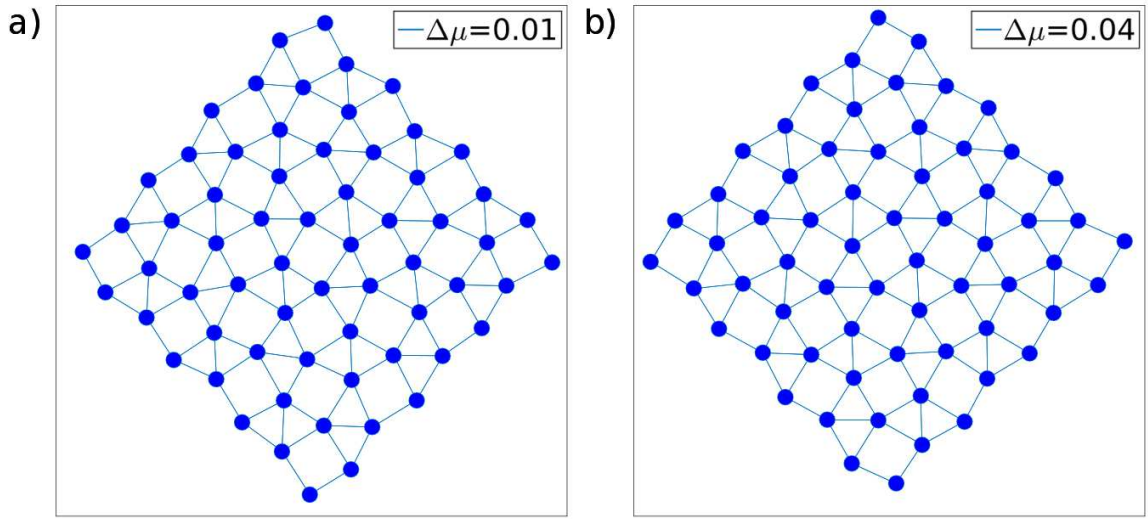}
	\caption{a) Configuration snapshots from Monte Carlo simulations of equilibrium snub square lattices that self assembled from the fluid upon quenching to $T=0.0309$ at $\rho=1.425$. Two cases, for pair potentials obtained via optimizations with chemical potential advantage constraints (a) $\Delta\mu=0.01$  and (b) $\Delta\mu=0.04$ of the target structure over select `flag-point' competitors are shown and discussed in the text. See Section ~\ref{sec:Methods} for simulation details.} 
	\label{fig:snub_conf}
	\end{figure*} 
 	 
    As detailed in Section \ref{sec:Methods},  Monte Carlo simulations of particles interacting with these optimized pair potentials were able to assemble into the snub square lattice from the higher-temperature fluid. While larger systems displayed sluggish assembly kinetics, smaller systems readily assembled into the the designed structure. When compared at the same temperature, potentials designed with $\Delta\mu=0.01$ and $\Delta\mu=0.04$ showed structural differences that point to enhanced thermal stability of the latter. As shown in Figure~\ref{fig:snub_conf}, assembled lattices of the potential designed with the stronger constraint (panel b) displayed less irregularity in the characteristic triangular and square tiling motifs than those with the weaker constraint (panel a). 

	\begin{figure}[ht]
	\includegraphics[scale=0.27]{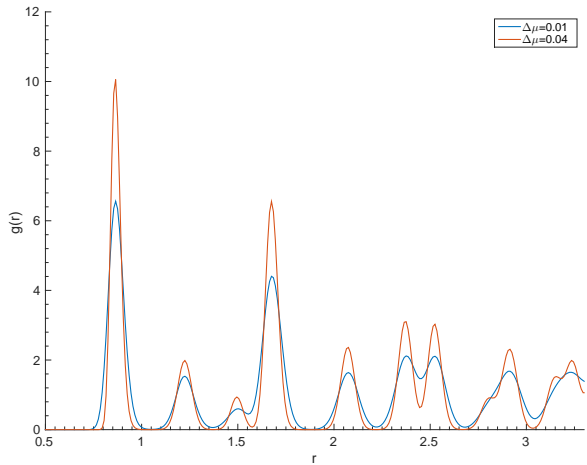}
	\caption{Radial distribution functions of equilateral snub square lattices that assembled from a fluid quench at $T=0.0309$. The two cases shown are with pair potentials obtained via optimizations with chemical potential advantage constraints $\Delta\mu=0.01$ (blue) and $\Delta\mu=0.04$ (red) of the target structure over select `flag-point' competitors. Both distributions have excellent agreement with those of their respective 
	perfect equilibrium snub square crystals (not shown).} 
	\label{fig:snub_gr}
	\end{figure}
    These differences are better appreciated in the radial distribution functions for both potentials where significantly sharper resolution of coordination shells is achieved for the $\Delta\mu=0.04$ potential as can be seen in Figure~\ref{fig:snub_gr}. This is poignant in the difficult to stabilize third shell of the snub square lattice where a {\em single} neighbor particle is expected to reside. This third shell neighbor corresponds to the third peak in the radial distribution, which as seen, is broader and somewhat  overlapped with that of the fourth peak for the $\Delta\mu=0.01$ potential, whereas it is well resolved for the $\Delta\mu=0.04$ case.  A similar trend likewise holds for longer range order ($r \sim 3$) where neighboring peaks coalesce for the $\Delta\mu=0.01$ system, while they remain distinguishable for the $\Delta\mu=0.04$ potential. Of course, peak resolution improves at lower temperatures for both potentials, but that this distinction can be observed for the higher temperature presented here ($T=0.0309$) is a testament of the superior thermal stability obtained by the interaction designed with a higher $\Delta\mu$ constraint. Another indication is (naturally) the melting temperature itself, where the first melt temperature exhibited by the $\Delta\mu=0.04$ interaction in Monte Carlo simulations was $23\%$ higher than for the $\Delta\mu=0.01$ potential for the same density (not shown). As expected from the earlier work on designing square lattices \cite{SquLat_dmu_opt}, however, the trade off of this improved thermal stability was the substantially reduced density range of stability for the ground state $\Delta\rho$, which was over four times lower for the $\Delta\mu=0.04$ system as compared to the $\Delta\mu=0.01$ potential (0.05 vs 0.23 respectively).

\section{Conclusion}
\label{sec:Conclusion}
	We have extended our recently introduced inverse design approach to find, to our knowledge, the first isotropic, convex repulsive pair potentials that favor assembly of two contrasting two-dimensional structures:  the kagome and the equilateral snub square lattice.  
	
	The kagome lattice's symmetric distribution of neighbors in its coordination shells make it particularly amenable to stabilization by an isotropic potential. For this structure, we use our optimization framework to design a strict-convex repulsive pair potential that maximizes the density for which the target is the ground state. We find that particles interacting via this potential readily self assemble into a kagome lattice from the fluid state upon cooling in Monte Carlo simulations.  
	
     On the other hand, designing isotropic, convex repulsive pair potentials that favor the equilateral snub square lattice is more challenging due to the asymmetric distribution of neighbors in its coordination shells and the presence of several closely related competing structures. To help address these challenges, and thereby provide a significant free energy gap between the snub square lattice and competitors \cite{SquLat_dmu_opt}, we required that the target ground-state structure maintain a minimum chemical potential advantage $\Delta \mu$ over select competitors during the parameter optimization. We find that, while larger systems of particles display sluggish assembly kinetics, smaller systems of particles interacting with the optimized potentials readily assemble into the snub square lattice from the fluid. As expected, based on previous work designing potentials for square lattices \cite{SquLat_dmu_opt}, a stronger $\Delta \mu$ constraint in the optimization led to enhanced thermal stability of the resulting snub square lattice but a significantly reduced density range for which it was the stable ground state.  
     
\begin{acknowledgments}
We acknowledge Prof. Jeffrey Errington for helpful conversations regarding the Monte Carlo quench simulations. 
T.M.T. acknowledges support of the Welch Foundation (F-1696) and the National Science. 
Foundation (CBET-1403768). We also acknowledge the Texas Advanced Computing Center (TACC) at the University of Texas at Austin for providing computing resources used to obtain results presented in this paper.
\end{acknowledgments}
\appendix* 	
\section{Free Parametrization of n-Basis Crystals for Forward Crystal Discovery}
	In carrying out the forward calculation of the ground-state structures from a pair potential described in Section \ref{sec:CompLattices}, one must determine the optimal values of any possible variable parameters for the lattices that minimize their chemical potential. For example, in the case of a Bravais lattice, one finds optimal values of aspect ratio $b/a$ and primitive vector angle $\theta$. Here, we generalize this and consider free variable primitive vectors with multiple freely parametrized basis. 
	
	Concretely, we consider a variable primitive lattice cell with [$b/a$, $\theta$] and include freely varying basis parameters [$b_{i}/a$,$\theta_{i}$] where $i$ indicates the basis number. Further, these basis are constrained to avoid superposition (i.e. distances between bases $>0$ ) and to remain within the primitive cell (e.g. $b_{i}/a\times\sin{\theta_{i}} < b/a\times\sin{\theta}$ etc). Physical quantities (energy, pressure etc) are then computed using standard formulation and the particle configuration that gives the lowest chemical potential found. We note a similar procedure was also used by Batten et al for a comparable forward calculation \cite{BT_QuinticPotentialKagomeHoneycomb}.
	\begin{figure}[ht]
		\includegraphics[scale=0.3]{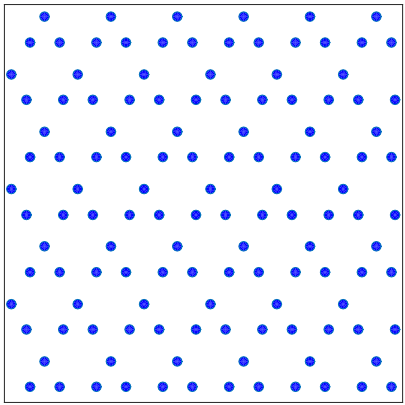}
		\caption{Illustration of the `Kagome-B' competing structure.}
		\label{fig:kagomeB}
	\end{figure}

	 Using this formulation the additional, non-standard lattices mentioned in section \ref{sec:CompLattices} were found. Specifically, for the kagome design target this resulted in a distorted honeycomb (2 basis: 1 primitive, 1 additional) with parameters [$b/a=1.114, \theta=1.117, b_{1}/a=0.877, \theta_{1}=0.663$]; and `kagome-B', a kagome lattice with non-uniform aspect ratio (3 basis: 1 primitive, 2 additional--see figure \ref{fig:kagomeB} for an illustration) with parameters [$b/a=1, \theta=\pi/3, b_{1}/a=0.447, \theta_{1}=0,b_2/a=0.447, \theta_2=\pi/3$] . Similarly for snub square, a distorted snub square was found (4 basis: 1 primitive, 3 additional) with parameters [$b/a=1, \theta=\pi/2, b_{1}/a=0.518, \theta_{1}=1.348, b_{2}/a=0.506, \theta_{2}=0.223, b_{3}/a=0.865 , \theta_{3}=0.796$]. 

%
\end{document}